\documentclass[aps,pra,twocolumn]{revtex4-1}
\usepackage{amsmath,amsfonts,graphicx}
\usepackage{amssymb}
\usepackage{epsfig,graphicx}
\usepackage{mathrsfs}
\usepackage[active]{srcltx}
\usepackage{bbm}

\newcommand{\spc}[1]{\mathcal{#1}}

\def\d{{\rm d}}

\def\>{\rangle}
\def\<{\langle}

\newcommand{\st}[1]{\mathbf{#1}}
\newcommand{\bs}[1]{\boldsymbol{#1}}     

\newcommand{\map}[1]{\mathcal{#1}}
\newcommand{\Tr}{\operatorname{Tr}}

\begin{document}

\title
{Superactivation of  quantum gyroscopes}
\author{Giulio Chiribella, Rui Chao, and Yuxiang Yang} \affiliation{Center for Quantum Information, Institute for Interdisciplinary Information Sciences, Tsinghua University, Beijing 100084, China}
\begin{abstract}
Quantum particles with spin are the most elementary gyroscopes existing in nature.  
Can  two such gyroscopes help two distant observers 
find out their relative orientation in space?
Here we 
show that a  single pair of gyroscopes in an EPR   state   gives little clue about the relative orientation,
but   when two or more EPR pairs are  used in parallel, suddenly a common reference frame  emerges, with an error that drops quickly  with the size of the system, beating  than  the best classical scaling already for small number of copies.
 This activation phenomenon indicates the presence of a latent resource  hidden into EPR correlations, which can be unlocked and turned into advantage when multiple copies are available.

\end{abstract}

\maketitle

\section{Introduction}

Reference frames play a central role  both in physics and everyday life. Simple words  like ``up" and ``down" or ``left" and ``right" only make sense relative to a  reference frame of spatial directions, like the direction of the gravitational field on the surface of the Earth, the frame of the fixed stars, or  the orientation of a gyroscope.
  In order to have a meaningful conversation,  two parties who are referring to   spatial directions need to share  the same reference frame, or at least to have their reference frames correlated in a precise way known to both.
       This is usually not a problem, because  the physical systems used as reference are large and classical, allowing  one to identify spatial directions up to a  negligible error.  However,
    deep down at the quantum scale the situation is radically different \cite{FirstRef,RMP}.   First of all, quantum particles are  tricky indicators of direction   \cite{holevo,massarpopescu,gisinpopescu}   and a well-defined reference frame can emerge from them only in the macroscopic limit \cite{peresscudo,bagan,prl2004,bagan2004}.   Second, quantum particles can be correlated in  strange new ways
    that were impossible  in  classical physics---ways
    that  puzzled Einstein, Podolski and Rosen (EPR) \cite{EPR} and led to the formulation of Bell's theorem \cite{Bell}.  When two particles are in an EPR state, each of their spins  does not possess  an individual orientation in space. The latter  pops up only at the moment of a measurement, in a manner that Einstein  skeptically dubbed
     ``spooky action at distance".
    How can two parties extract a common reference frame  from such spooky  quantum
 correlations?


The answer turns out to be  surprising. 
Think of the most elementary gyroscope,  a spin-$j$ particle, and  imagine that two parties are  given two gyroscopes in an EPR  state.    Relying on the correspondence principle \cite{Bohr},  one could expect that a well-defined  reference frame emerges in the limit of large spin.  But in fact this is not the case:  here we show that, no matter how large $j$ is, there is always a residual    error.
   If the two spins were classical gyroscopes, this would mean that their directions are not sufficiently well correlated to establish a common reference frame reliably.    In spite of this intuition too, when two or more EPR pairs are available we find that  suddenly a well-defined reference frame emerges.    The error vanishes at the classical rate $1/j$ for two pairs and achieves the  non-classical  scaling $\log j/j^2$ for three pairs.  From four pairs  onwards the two parties can establish a common reference frame with Heisenberg-limited error of order $1/j^2$, the ultimate scaling compatible with  the size of the gyroscopes.
 This result spotlights the presence of a latent resource,  contained into EPR correlations, which is activated when multiple copies are observed jointly.




\section{Results}

\noindent {\bf Remote alignment of reference frames.}
 Consider the following  Gedankenxperiment.   Two pilots,  Alice and Bob,  are about to leave a ground station for a  mission at two distant satellite stations.    In preparation for the trip,  they correlate their gyroscopes---for example, by aligning them with the axes of the ground station.
Then, they take off for their respective destinations and during the journey their   local reference frames become misaligned,   undergoing to two unknown rotations $g_A$ and $g_B$. Once arrived,  Alice and Bob need  to realign their
 reference frames,  performing on their axes two rotations $h_A$ and $h_B$ such that $h_A  g_A  =   h_B  g_B$.
 At this stage, only the exchange of classical messages  is possible,~e.~g.~via radio signals.  Hence,  Alice and Bob can only rely on the correlations between their gyroscopes, established when they were still at the ground station \cite{Note1}.
 
In general, a perfect alignment cannot be achieved  with gyroscopes of finite size.    A natural  way to quantify the error is to consider the square distance between Alice's and Bob's axes after the realignment, averaged over the three axes \cite{holevo,peresscudo,bagan,prl2004,spekkensrmp}. Explicitly, the error is given by
  \begin{align}\label{error}
 d^2(h_A,h_B,g_A,g_B)    &=   \frac 13  \sum_{i=x,y,z}   \left  \|   h_A g_A \st n_i -   h_B   g_B   \st n_i  \right \|^2  \, ,
  \end{align}
 where $\st n_x ,\st n_y$ and $\st n_z$  the unit vectors pointing in the directions of the $x,y$ and $z$ axes at the ground station, respectively.    The goal of the alignment protocol is to minimize this error, making the best possible use of the correlations between Alice's and Bob's gyroscopes.



\begin{figure}
  \centering
  \includegraphics[width=0.9\linewidth]{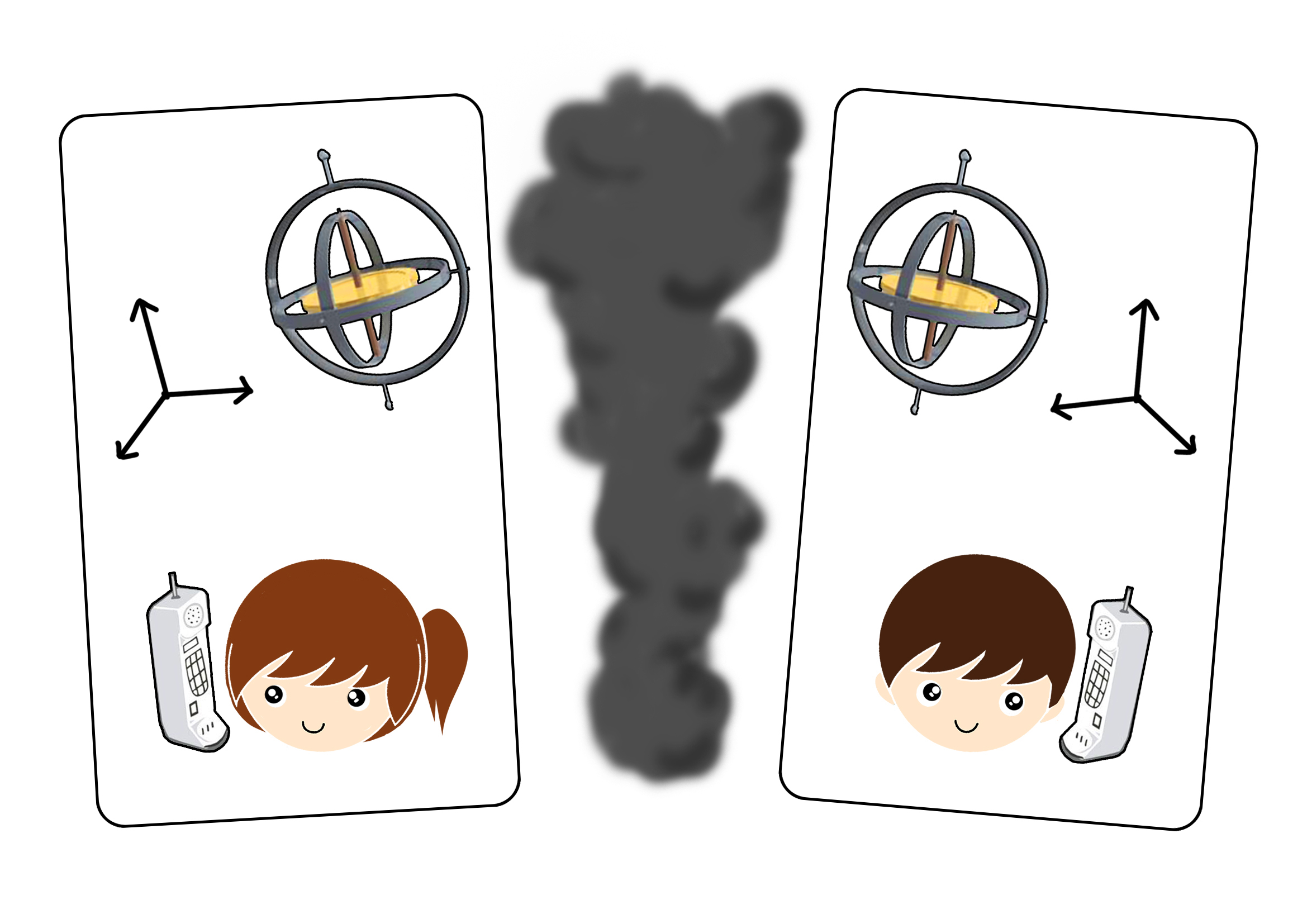}\\
  \caption{{\bf Remote alignment of reference frames.}  Alice and Bob, based in two distant satellite stations, perform local operations on their gyroscopes, coordinated by the exchange of classical messages. Their goal is to align the local reference frames  at the two stations, using the information extracted from   the correlations between the gyroscopes.}\label{fig:loccalign}
  \end{figure}

Let us look more closely into the problem.    Suppose that Alice's and Bob's gyroscopes are quantum systems with Hilbert spaces $\spc H_A$ and $\spc H_B$, respectively, initially prepared in the state $\rho_{AB}$.  For simplicity, we choose  $\spc H_A  \simeq \spc H_B$.    The alignment protocol consists of local operations (LO),  performed at the satellite stations, coordinated by classical communication (CC) between the two stations, as in Figure \ref{fig:loccalign}.  Eventually, the protocol outputs the classical description of the two rotations  $h_A$ and $h_B$.  All together,  this  is described by an \emph{LOCC measurement}, performed on the two gyroscopes, with outcomes $h_A$ and $h_B$.  We denote the measurement operators by $M_{h_A,h_B}$.   Here it is crucial to recall that Alice and Bob perform  operations relative to their local reference frames: from the point of view of the ground station, the actual measurement  is described by the  operators
$M^{g_a,g_B}_{h_A,h_B} =   (\map U^{-1}_{g_A}  \otimes \map U^{-1}_{g_B})  \left(M_{h_A,h_B}\right)$,
where  $\map U_{g}  (\cdot)  :=  U_g\cdot U_g^\dag$ is the unitary channel representing the rotation $g$ on the Hilbert space of a single gyroscope.
The goal of the measurement is to minimize the expected  error, in the worst case scenario over all possible rotations $g_A$ and $g_B$.
The usefulness of a state $\rho_{AB}$ for Alice's and Bob's  task is then quantified by the minimum value of the error over all possible measurements.

\medskip

\noindent{\bf   No reliable alignment from  a single EPR pair.}
 Our first result is that the EPR correlations between two spin-$j$ particles  are not sufficient to establish a reference frame reliably.  Suppose that  the two particles are in the singlet state
 \begin{align}\label{singlet}
  |S_j\>   =   \frac 1 {\sqrt{ 2 j +1}}   \sum_{m=-j}^j  (-1)^m  |j,m\>  |j,-m\> \, ,
  \end{align}
  where $|j,m\>$ is the eigenstate of the $z$-component of the angular momentum operator \cite{peresbook}.
  In this case we prove that the alignment error  is lower bounded as   \begin{align}
 \label{lowerbound}   \left\<  d^2  \right\>\ge \frac 43
 \end{align}
 \emph{independently of the size of the spin} (cf. Methods).    The value $4/3$ is the  error that would be found if Alice and Bob managed to align perfectly their $z$ axes, but ended up with the $x$ and $y$ axes  rotated by a completely random angle.
Furthermore, we show that the lower bound $\left\< d^2\right\> \ge 4/3$ can be attained if,  in addition to the two spin-$j$ particles, Alice and Bob have an EPR pair of  $(2j+1)$-dimensional quantum systems that are invariant under rotation.  These   systems  can be realized~e.~g.~by the charge or current states of a solid state quantum device, or by a virtual subsystem of a set of spin-$1/2$ particles \cite{spekkensrmp}.       Taking into account of this extra resource, the joint state of Alice's and Bob's systems is then given by
\begin{align}\label{state}\rho_{AB}  =  |S_j\>\<S_j|_{A_1B_1}\otimes |\Phi^+_j\>\<\Phi^+_j|_{A_2B_2} \, ,
\end{align}
where    $A_1, B_1$  label the two spin-$j$ particles, $A_2,B_2$ label the auxiliary systems, and $|\Phi^+_j\>$  is the standard EPR  state in dimension $2j+1$.
   Now, thanks to the assistance of the  irrotational EPR  pair, Bob can use the quantum  teleportation protocol \cite{teleportation} to  transfer the state of his spin-$j$ particle to Alice, up to the rotation that Alice needs to invert in order to align her axes  (cf. Methods). Estimating this rotation and performing the corresponding correction, Alice can achieve  the  minimum error $\left\< d^2\right\>= 4/3$.
Remarkably, this value is achieved thanks to the assistance of quantum correlations between systems that carry no information  whatsoever about directions.  We will come back to this point later in the paper.  



\medskip

\noindent{\bf  Weak activation   with two EPR  pairs of spin-$j$ particles.}
For a single copy of the spin-$j$ singlet  we have seen that the alignment error is independent of $j$.
The  situation suddenly changes when  if two copies are available: in this case, we show that the alignment error vanishes with standard quantum limit (SQL) scaling $1/j$.
For large $j$, we show that the error is lower bounded as  $\left\<d^2\right\>  \ge  2/(3j)   +   O(  \log j/j^2)$ (cf. Supplementary Note 1).
 Again,  this bound  can be attained  using the  teleportation trick, provided that Alice and Bob are assisted by EPR correlations among rotationally invariant systems.  

\begin{figure}
  \centering
  \includegraphics[width=0.9\linewidth]{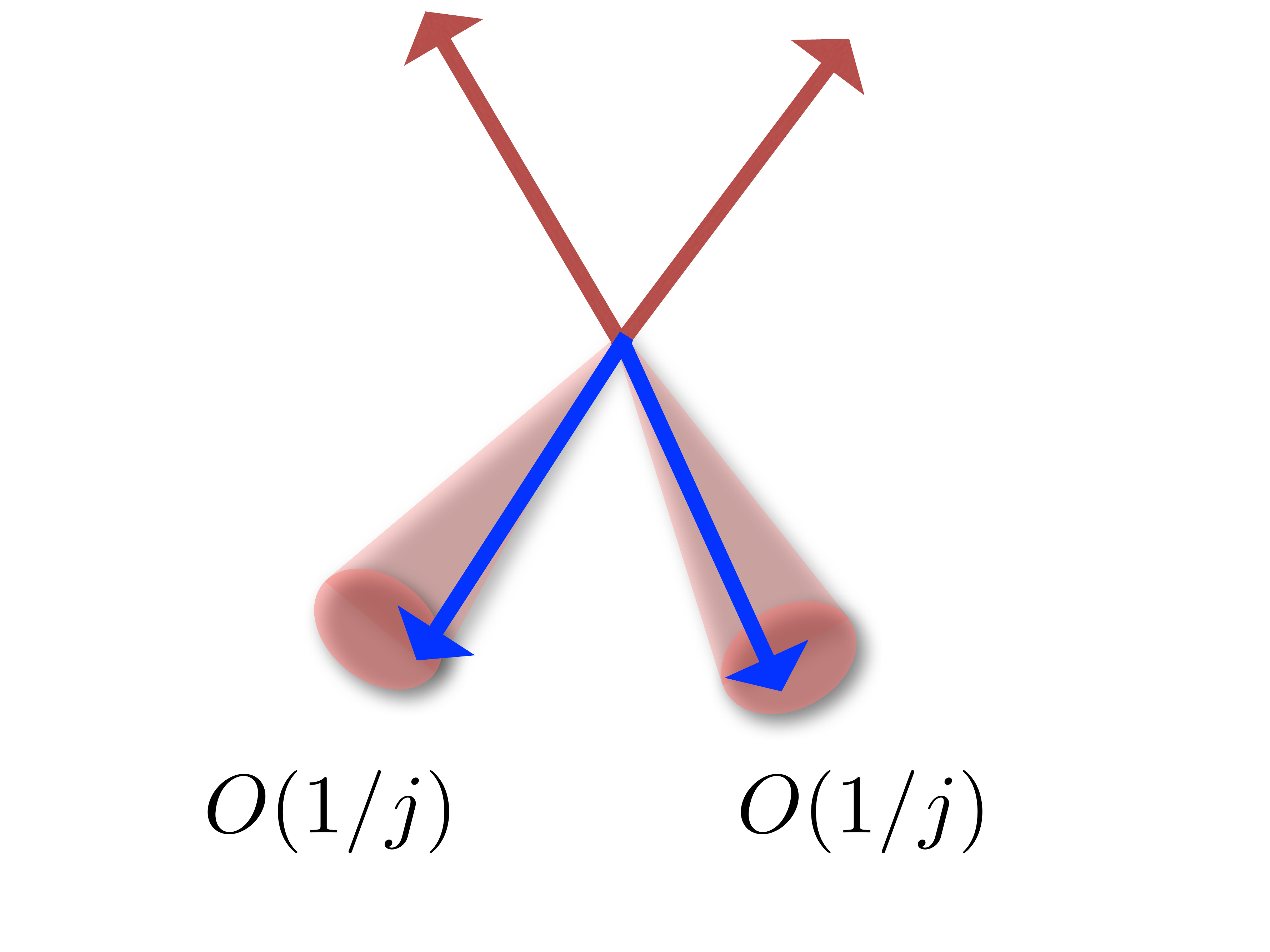}\\
  \caption{{\bf Classical model for weak activation.}
  Alice and Bob use two pairs of classical gyroscopes, with Alice's (Bob's) gyroscopes represented by red (blue) arrows. For each pair, Alice's gyroscope  points in a random direction and Bob's gyroscope  is within a solid angle of size $O(1/j)$   (shaded red cone) around  the opposite direction.  When two such pairs are used, with high probability the two gyroscopes on Alice's side will point in two distinct directions, thus identifying a Cartesian reference frame with square error $O(1/j)$.
    }\label{fig:classical}
  \end{figure}

The reduction of the error for two EPR pairs is a weak form of  activation: Alice and Bob can align their axes by combining two resources that individually  did not allow for alignment.
However, this phenomenon  is not very surprising  {\em per se}, because it can be reproduced  by a simple classical model, illustrated in Figure \ref{fig:classical}.
Suppose that two classical gyroscopes are prepared in the following way: Alice's gyroscope points in a random direction in space and Bob's gyroscope points in the opposite direction, up to an error (square distance) of size $1/j$.     Using a single pair of gyroscopes, Alice and Bob can align at most one axis, so their error will be at least  $4/3  +  O(1/j)$. However, if Alice and Bob have two pairs of gyroscopes, then with high probability they will be able to identify two distinct directions in space, up to an error of order $1/j$.  Clearly, once they have established two directions, they will be in position to reconstruct a full reference frame, using~e.~g. the right-hand rule \cite{Note2}.  In other words, the emergence of a reference frame with error $1/j$ is still compatible with a hidden variable model where the spins have definite orientation prior to the measurement.

\medskip
\noindent{\bf Superactivation of quantum sensitivity. }       The existence of a classical explanation for the scaling $1/j$ may look reassuring, but this superficial impression is deceitful:  when it comes to indicating directions in space, two EPR pairs  contain much more than it  first meets the eye.  Here we show that using a suitable quantum measurement, Alice and Bob  have the chance  to  extract a reference frame with error vanishing  with Heisenberg limit  (HL) scaling $1/j^2$---a scaling that would not be possible if the individual orientation of the particles were a real property  defined prior to the measurement.

To  see how this phenomenon arises,  we decompose the product of two spin-$j$ singlets  as
  \begin{align}\label{state2}
   |S_j\> |S_j\>  =   \bigoplus_{k=0}^{2j}   \sqrt{p_k}  \,   |S_k\>  \qquad   p_k   =   (2k+1)/(2j+1)^2  \, ,
   \end{align}
   where  $|S_k\>$ is the  spin-$k$ singlet contained in the tensor product $\spc H_{k,A}  \otimes \spc H_{k,B}$, with  $\spc H_{k,A}$  ($\spc H_{k,B}$) being the subspace of  $\spc H_{A_1}\otimes \spc H_{A_1}$  ($\spc H_{B_1}\otimes \spc H_{B_1}$) with total angular momentum  number equal to $k$.
 Now, Alice and Bob  can apply  a protocol that separates two branches of the wave function,  with the feature that in one branch the error vanishes with Heisenberg limit  (HL) scaling $1/j^2$, while in the other branch the scaling is still $1/j$.   The two branches are   separated by a filter with operators $\{F_{\rm yes}, F_{\rm no} \}$, so that, if the outcome of the filter is $x $, the state of the gyroscopes is $\rho_{AB,x}   =    F_x  \left ( \rho_{AB}  \otimes \rho_{AB}  \right)  F_x^\dag/   p_{x}$, where $p_x  =   \Tr[F_x  \left ( \rho_{AB}  \otimes \rho_{AB}  \right)   F_x^\dag]$ is the probability that the filter heralds the outcome $x$.

  Let us see how much the error can be reduced in the favorable branch. First, using the teleportation trick (cf. Methods), Bob can transfer his part of the singlets to Alice, who ends up holding two copies of the rotated singlet state $|S_{j,g}\> :  =   (  U_g  \otimes I)  |S_j\>$.
  Crucially, the state  $   |S_{j,g}\> |S_{j,g}\>$ has the same form of the optimal state for the transmission of a Cartesian frame \cite{prl2004,bagan2004,hayashi2006pla}, with the only difference that the coefficients of the latter are given by
  $p_k^{\rm opt}  =   \sin^2\left[\frac{\pi(k+1)}{2(j+1)}\right]/(j+1)$.
  Hence, choosing  a filter that remodulates the coefficients of the wavefunction,  the  initial state  can be transformed into the optimal one, thus reducing the error to  $\left\<d^2\right\>  =  \pi^2/(6j^2)  +  O  (1/j^3)$, the absolute minimum set by quantum mechanics for composite systems of angular momentum upper bounded by $2j$   \cite{pra2005}.

  Let us see how large  is the probability of this precision enhancement.
    First of all,   the filter that activates the Heisenberg scaling  can be achieved even before the teleportation step, using a single local operation, say, in Bob's laboratory.  The desired modulation is achieved if Bob filters the state of his spin-$j$ particles with the operator $F_{\rm yes,  B}   =   \lambda  \,  \sum_k   c_k^{\rm opt}/c_k \,  P_{k,B}$, where $\lambda >  0$ is a suitable constant and $P_{k,B}$ is the projector on $\spc H_{k,B}$.
     \begin{figure}
  \centering
  \includegraphics[width=0.9\linewidth]{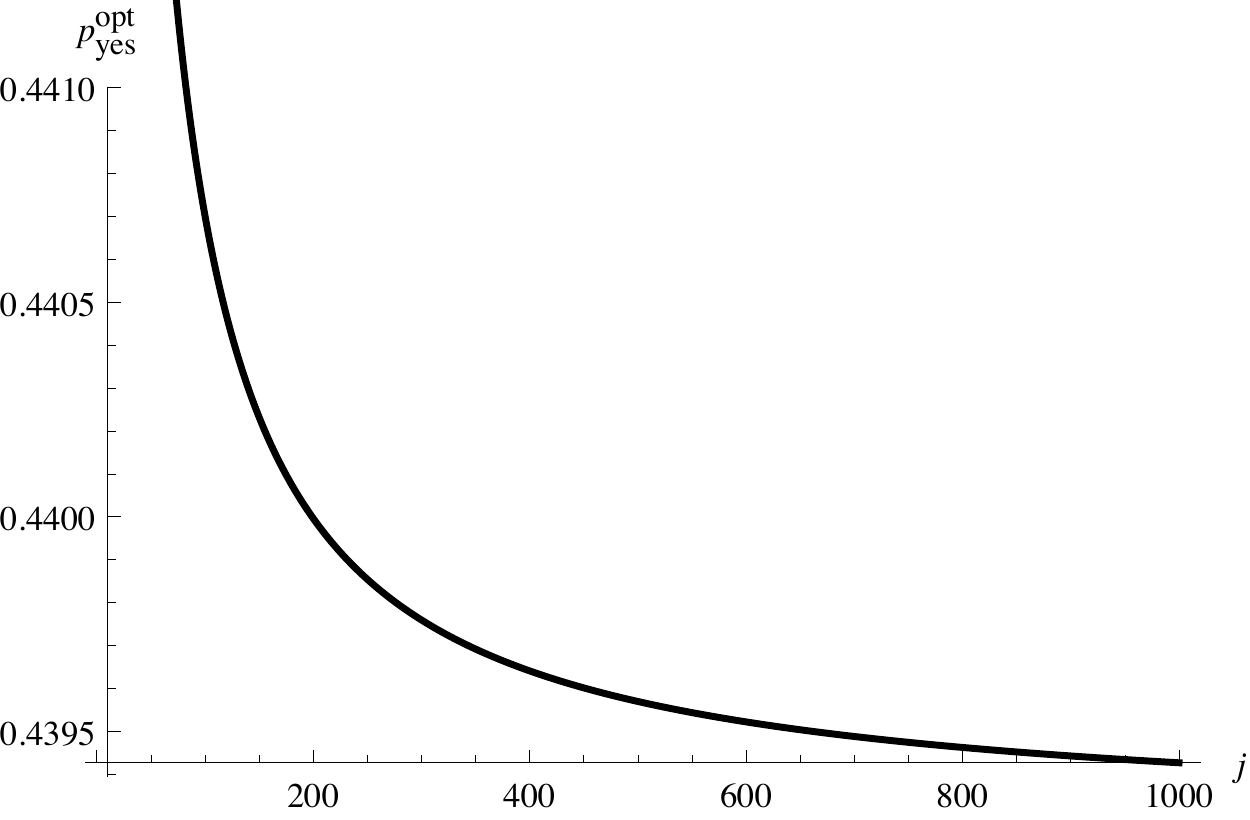}\\
  \caption{{\bf Probability of achieving  alignment at the Heisenberg  limit with two  EPR pairs.}      The optimal value of the probability is plotted for  $j$ varying from $10$ to 1000 in unit steps.  For small  $j$ the value is always larger than $44\%$, while for $j\to \infty$ it converges to $43.93\%$. }\label{pyes}
  \end{figure}
Since the filter operator $F_{\rm yes, B}$ must be a contraction, we have the achievable upper bound $p_{\rm yes}  =  \lambda^2    \le \min_k c_k^2/\left(c_k^{\rm opt}\right)^2$. Hence, the maximum probability of the favourable outcome is given by the expression
  \begin{eqnarray}
   p_{ \rm yes }^{ \rm opt }
   &=& \min_{k } \frac { (2k+1) (j+1) } { (2j+1)^2
   \sin\left[ \frac{\pi(k+1)}{2(j+1)} \right]^2 } \, ,
  \end{eqnarray}
 which converges to $43.9\%$ in the large $j$ limit. The exact values of $p^{\rm opt}_{\rm yes}$ for $j$ up to 1000 are shown in Fig. (\ref{pyes}), note that the value is above  $43.9\%$ for every value of $j$.

 One may ask what  happens in the remaining 56.1\% of the cases, when the filter gives the unfavourable  outcome.  Is the error still scaling with $j$?  And, if yes, how?      Luckily, we find that in these cases the error maintains the SQL scaling  $\left\<d^2\right\>  \approx  1.189/j$ (cf.  Supplementary Note 2), with a constant that  is less than twice the constant appearing in  the optimal deterministic strategy. 

In summary, we have seen that two EPR pairs of spin-$j$ particles allow Alice and Bob to align their axes up to an error scaling like $1/j^2$ in at least $43.9\%$ of the cases.  This scaling  is incompatible with the assumption that the four particles used by Alice and Bob have a definite orientation in space.  Indeed, 
 a  single spin-$j$ particle cannot indicate a direction with error smaller than $O(1/j)$ \cite{holevo,massarpopescu}.  This implies that, if each particle had a definite orientation, then the error using four particles would still vanish as $O(1/j)$ \emph{for a single direction}---not to speak about  a full reference frame.   In summary, the activation of the Heisenberg scaling  highlighted here  is  radically different from the weak activation that one can see in the classical world. Essentially,   it  is based on  the fact that the EPR particles do not have any orientation prior to the measurement, and when two EPR pairs are available, the particles can be steered into the most sensitive state possible.    
 
  We refer to  \emph{superactivation  of quantum sensitivity} whenever the error vanishes faster than the classical scaling $1/j$.
 It is important to stress that this phenomenon 
  is not an artifact of the specific error function used in our calculation:  as a matter of fact, superactivation occurs generically for the expectation value of every  bounded  cost  function  $f  (h,g)$ that reaches its absolute  minimum  $f_{\min}$ only when   the axes are aligned  ($h=  g^{-1}$) and admits a second-order Taylor expansion  in a neighbourhood of the absolute minimum. For example,  superactivation occurs for the variance of the three Euler angles.
The easiest way to see this is to note that, by Chebyshev's inequality,  the probability that after the execution of the protocol  the distance  between Alice's and Bob's $i$-axis,  $i=  x,y,z$, is larger than $\epsilon$ 
 is upper bounded as
\begin{align}     {\rm Prob}  [ d_i  >  \epsilon  ]   \le       \left\<d^2 \right\>/\epsilon^2 \, .
\end{align}

  By Taylor expansion,  this implies that $\<f\>$ has to tend to the minimum value $f_{\min}$ as fast as $\left  \<   d^2\right\>$ tends to zero---in particular, it has to tend to $f_{\min}$  as $1/j^2$ when the filter gives the favourable outcome.              On the other hand, for a single copy the average cost must remain bounded away from $f_{\min}$:  otherwise, the probability that Alice's and Bob's axes are misaligned should vanish, and so  should do the error $\left\<d^2\right\>$, in contradiction with Eq. (\ref{lowerbound}).


\medskip

\begin{figure}
  \centering
  \includegraphics[width=0.9\linewidth]{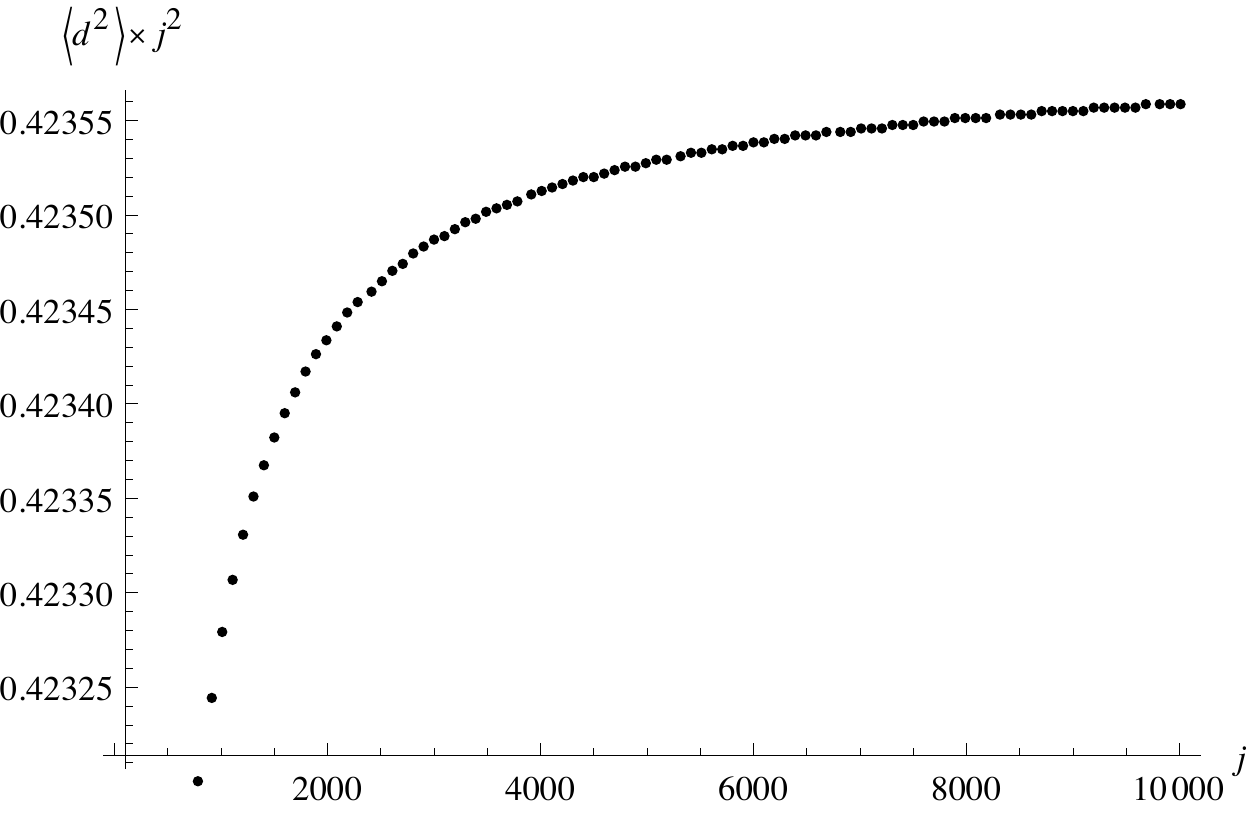}\\
  \caption{{\bf Heisenberg limited alignment with  four EPR pairs.}   The exact value of the alignment error, multiplied by $j^2$, is shown here for  $j$ ranging from  100 to 10000 in steps of 100.  For large $j$ the plot exhibits the Heisenberg scaling  $\left\<d^2\right\>  =11\ln(2)/(18j^2)$, in agreement with our analytical result.  }\label{four}
\end{figure}

\noindent{\bf Deterministic  superactivation.}
 The  probability of reaching the HL can be further amplified by repetition of the protocol, which allows one to attain HL precision with probability $p_n  > 1-   (0.561)^n$ using $2n$ EPR pairs.  However,  one can do even better:   taking advantage of joint measurements, the HL can be achieved \emph{with  certainty}  using only four EPR pairs.  To establish this result, we observe that the  state $|S_j\>^{\otimes 4}$ can be viewed as a quantum superposition  of spin-$k$ singlets as in Eq. (\ref{state2}), with the difference that now $k$ ranges from 0 to $4j$ and the coefficients $c_k$ have a different expression (cf. Supplementary Note 3).  Using this fact, we show that the error scales as
 \begin{align}
 \left\<d^2\right\>= \frac{11\ln 2}{18j^2}    +  O(j^{-3})     \, ,
 \end{align} proving that four copies are sufficient to attain the HL with a deterministic strategy.  The exact values of the error are shown  in Figure \ref{four} for $j$ up to $10000$.

\begin{figure}
  \centering
  \includegraphics[width=0.9\linewidth]{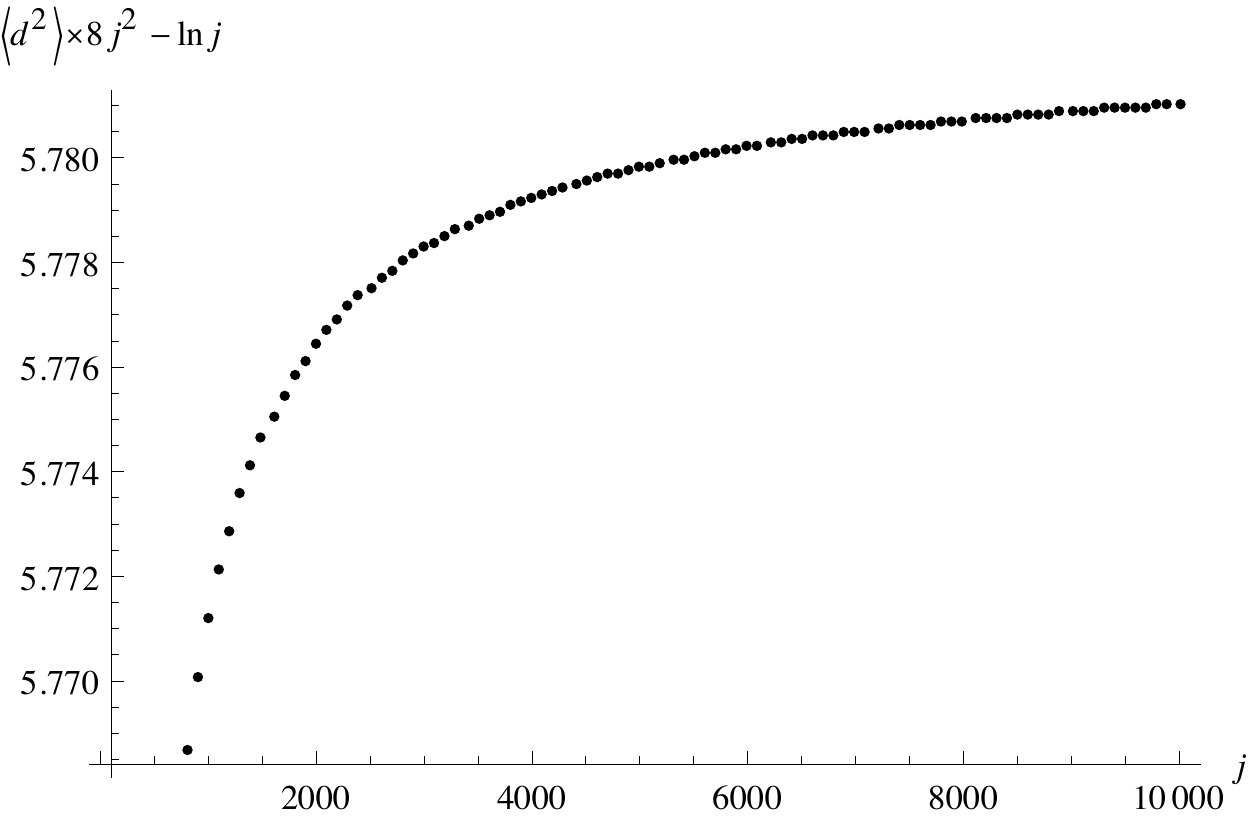}\\
  \caption{{\bf Quasi-Heisenberg scaling of the alignment error with three EPR pairs.}   The plot shows the exact values  of the function  $ 8j^2 \left\<d^2\right\>  -\ln j $ for  $j$ going from 100 to 10000 in steps of 100. For large $j$ the plot exhibits the quasi-Heisenberg scaling $\left\<d^2  \right\>=  \ln (j)/(8j^2)$, 
  again in agreement with our analytical result. }\label{three}
\end{figure}

Interestingly, four copies are strictly necessary to achieve the HL with unit probability. Nevertheless, with three copies Alice and Bob can  still achieve superactivation, reducing the error to the quasi-Heisenberg scaling $\left\<d^2\right\>=  \ln (j)/(8j^2)  +  O(1/j^2)$ (cf. Supplementary Note 4). The exact values of the error are plotted in Figure \ref{three}.

\medskip

\noindent{\bf  Quantum metrology with  spin-$j$ singlets.}
In the previous paragraphs we presented our results in a bipartite communication scenario. However, using the technique shown in Methods, it is immediate to translate them into the conventional single-party scenario of quantum metrology \cite{MetroRev1,MetroRev2}. In this formulation, the problem is to estimate an unknown rotation  $g$ from $n$ copies of the rotated spin-$j$ singlet  $   |S_{j,g}\>$.   This problem arises e.g. in high precision magnetometry \cite{magnetometry, PRLmagnet,Noisymagnet}, for setups  that probe the magnetic field using  a spin-$j$ particle entangled with a reference \cite{EntMetro,PRAMetro}, or setups   designed to measure the magnetic field gradient between two locations \cite{mag1,mag2}.
In this scenario, the fact that quantum-enhanced precision can be achieved using  $n\ge 3$ spin-$j$ singlets is  good news, since  spin-$j$ singlets are much easier to produce than the optimal quantum states for the estimation of rotations  \cite{prl2004,bagan2004,hayashi2006pla}.  A concrete setup that  generates a spin-$j$ singlet using two spatially separated Bose-Einstein condensates of    $\phantom{}^{87}{\rm Rb}$ atoms was put forward in  Ref. \cite{PRA87}.
Still, the implementation of the optimal quantum measurement remains as a challenge.   

\medskip

\noindent{\bf  Bridging the gap with the Quantum Cram\'er-Rao bound.}
A popular approach to quantum metrology is via the quantum Cram\'er-Rao  bound (CRB), which lower bounds the variance with the inverse of the quantum Fisher information  \cite{fisher1,Helstrom,holevo,Fujiwara}.   The bound is known to be achievable  in the asymptotic limit where a large number of identical copies are available \cite{gillmassar,MatsumotoQCR,HayashiBook}. Practically, however, the CRB is often invoked to discuss quantum advantages in the single-copy regime.    Our result provides a warning that such an extrapolation can sometimes lead to paradoxical results:  
For one copy of a spin-$j$ singlet, it is not hard to see that Fisher information grows like $j^2$  \cite{PRA87,QFI} (see also Supplementary Note 5). This means that, if the  CRB were achievable in a single shot, the  variance  in the estimation of the three Euler angles would have to vanish as $1/j^2$.
 But we know that this is not possible:  if the variance vanished with $j$---no matter how fast---then also the average of the error in Eq. (\ref{error})  would have to vanish, in contradiction with our result.   In short, this shows  that the CRB is not achievable with a single copy.   The non-achievability of the CRB in the single-copy regime was observed for  phase estimation in Ref.  \cite{hayashiCMP}, although in that case the CRB was still predicting the right asymptotic scaling---only,  with a constant that was smaller than the actual one.  In the case of spin-$j$ singlets the effect is more dramatic: even the  scaling with $j$  is unachievable for a single copy.
   In order to achieve the  CRB, one needs a sufficiently precise information about the true value, which can be obtained e.g. in the limit of asymptotically large number of copies \cite{gillmassar}.   The achievability of the CRB in the large copy limit  can be seen explicitly in our approach.   Denoting by $n$ the number of copies, we find that the optimal measurement has  error given by  $\left\<d^2\right\>   =  3/\left[2nj(j+1)\right]  $   up to a correction of order $ n^{-3/2}  j^{-3}$ or  $n^{-2}  j^{-2}$, depending on the relative size of $n$ and $j$ (Supplementary Note 6).    The measurement that minimizes the error also  achieves the CRB (Supplementary Note 7).   Most importantly for the CRB approach, the achievability of the bound requires  $n$ to be large, but not necessarily  large compared to $j^2$.

\medskip

\noindent{\bf Buying enhanced  sensitivity with  correlated quantum coins.}
In the remote alignment protocols of this paper,  Alice and Bob achieve the minimum error  by  using in tandem  two different resources: the correlation between their spins and the correlation between two degrees of freedom that are insensitive to rotations.    In the  classical world, any such protocol would look extravagant---for sure, having  a number of correlated random bits  does not help Alice and Bob align their axes!  But the situation is radically different in the quantum world, where  correlated quantum bits can make a difference in the  precision of alignment.  Consider the simplest case $j=1/2$ and suppose that Alice and Bob use only two correlated spins, without the assistance of a rotationally-invariant EPR pair. In this case, it can be proven that the error must be at least   $\left\<d^2\right\>= 16/9$ (cf. Supplementary Note 8), strictly larger than the value   $\left\< d^2\right\> =  4/3$ that can be achieved with the teleportation trick.  In other words, correlations that \emph{per se}  are  useless for the alignment of reference frames can become useful   in conjunction with correlations among rotating degrees of freedom.
 This result  is deeply linked to the tasks of entanglement swapping \cite{entswap} and dense coding \cite{densecod} and highlights the non-trivial interaction between the resource theory of entanglement \cite{EntangleRMP}  and the resource theory of reference frames \cite{resource1,resource4,resource5}.


\section{Discussion}

The superactivation effect suggests a way to delocalize the ability to align Cartesian frames over different parties, in the spirit of  quantum secret sharing \cite{secretsharing2,secretsharing1,bagansecretdirection}.    Imagine that, in order to accomplish a desired task, the two satellite stations $A$ and $B$ must  have their reference frames aligned with high precision.  At the two stations there are  two groups of parties, $\{A_1,\dots, A_n\}$  and $\{B_1,\dots, B_n\}$, with each pair of parties $(A_i, B_i)$ possessing a pair of spins in an EPR state along with additional quantum correlations in invariant degrees of freedom.       Now, our result guarantees that a single pair alone cannot achieve the task: at least two pair of parties  have to cooperate in order to  reduce the error down to zero.  Moreover, if the task requires the error to be of order $1/j^2$ (instead of $1/j$ or $\log j/j^2$), then at least four parties at each station have to cooperate. Compared with the state of the art \cite{bagansecretdirection},  our protocol offers a quadratic enhancement of precision,  allowing one to achieve the Heisenberg limit.    On the other hand, our secret sharing protocol has necessarily a low threshold (four cooperating parties can always establish a reference frame reliably).  A promising avenue of future research consists in combining the best of the two protocols, thus having a secret sharing scheme that achieves the Heisenberg limit with every desired  threshold $t\ge 4$.

\section{Methods}

\noindent{\bf Reduction to parameter estimation with shared reference frames.}
In order to evaluate  the error,  we reduce  the alignment problem to a simpler form.
 First, note that the error does not change if one replaces the  original protocol with a protocol where Bob communicates  $h_B$ to Alice, who  performs the rotation $h_B^{-1}h_A$. Assuming classical communication as a free resource, one can restrict  the minimization of the error to protocols involving only one rotation on Alice's side. Therefore, we will drop $h_B$ everywhere  and use the  notation $h_A  \equiv h$.   In addition,  the error does not change if one rotates the reference frame of the ground station.  This means that  we can always pretend that $g_B$ is the identity rotation, provided that we replace $g_A$ by $g_Ag_B^{-1}\equiv g$. Averaging over all possible rotations, the error can be expressed in the form
 \begin{align}\label{error2}
\left \<d^2 \right\> =  \max_g  \int   \d h \,     d^2(h,g)    \,  \Tr  [ M_h    (\map U_g\otimes \map I_B)  (\overline{\rho_{AB}}) ]  \, ,
 \end{align}
 where  $d(h,g)  :  =  d(h,e,g,e)$ and  $\overline {\rho_{AB}}  :   =   \int \d k  \,  ( \map U_k \otimes \map U_k)  (\rho_{AB})$,  $\d k$ being the normalized Haar measure over the rotation group.
Eq. (\ref{error2}) has an important conceptual implication: the fact  that Alice and Bob have different reference frames has disappeared from the problem---instead, what remains is  only the LOCC estimation of the rotation $g$ from  the state $ (\map U_g\otimes \map I_B)  (\overline{\rho_{AB}})$.  

\medskip

\noindent{\bf Lower bound from global measurement.}
 A lower bound on the error can be obtained by lifting the LOCC requirement in the previous paragraph.    When Alice and Bob share a spin-$j$ singlet, this means finding the best global measurement that identifies the rotation $g$ from the state   $|S_{j,g}\> $, $U_g^{(j)}$ being the unitary that represents the rotation $g$ on the spin-$j$ particle $A_1$.     The optimal measurement can be found using  the method of Ref. \cite{pra2005}, which in this case gives $M^{\rm {opt}}_h= (2j+1)^2  \,  |S_{j,h^{-1}}\>\<  S_{j,h^{-1}}|$.         Plugging the optimal measurement into the r.h.s. of Eq. (\ref{error2}) it is straightforward to obtain  the lower bound $\left\<d^2\right\>  \ge   4/3$ for every possible value $j>0$  (see e.g. \cite{hayashi2006pla,pra2005}).
 The lower bound coincides with the error that would be found if Alice and Bob succeeded in aligning perfectly the $z$-axis, but had the $x$ and $y$ axes rotated by a random angle: indeed, setting    $\st n_z^B  \equiv  \st  n_z^A$,  $\st n_x^{B}   =   \cos \theta  \,  \st n^A_x   +  \sin \theta \,   \st n^A_y$    and $\st n_y^{B}   =   \cos \theta  \,  \st n^A_y   -  \sin \theta \,   \st n^A_x$ one has
 \begin{align*}
 \left\<  d^2\right\> &  =    \int \frac{\d \theta}{2\pi}  \,  \left [   \frac 13  \sum_{x,y}    2(  1  -    \cos \theta) \right] =   \frac 43 \, .
 \end{align*}

\medskip

\noindent{\bf Achievability of the bound.}     The bound obtained by minimizing the error \ref{error2} over all global measurements  can be achieved if Bob can transfer his part of the state to Alice by LOCC.  This is the case when Alice and Bob share  EPR correlations among degrees of freedom that are invariant under rotations.    These additional EPR pairs do not pick the rotation in Eq. (\ref{error2}) and therefore can be used as a resource to implement the quantum teleportation protocol \cite{teleportation}.  In this way, Alice is in position to perform the optimal global measurement  on systems $A_1 $ and $B_1$.  Since teleportation is a LOCC protocol, the whole procedure describes a valid LOCC estimation strategy and, thanks to  the reduction of  Eq. (\ref{error2}), a valid alignment protocol.

\medskip

\noindent {\bf Acknowledgments.}  This work is supported by the National Basic Research Program of China (973) 2011CBA00300 (2011CBA00301),  by the National Natural Science Foundation of China through Grants  11350110207, 61033001,  and 61061130540,  by the Foundational Questions Institute through the Large Grant ``The fundamental principles of information dynamics",  and by the 1000 Youth Fellowship Program of China.
We are grateful to Xinhui  Yang for drawing Figure 1.

\appendix

 \section*{Supplementary Note 1} 

Let us start from a general observation.  Consider the problem of estimating a rotation $g$ from the signal state $  |\psi_g\> $ 
 of the form
\begin{align}\label{Astate}
|\psi_g\>  =  \sum_{k=0}^{2j}  c_k|S_{k,g}\> \, , \qquad c_k  \ge 0 \,  \forall k \in  \{0,\dots, 2j\}  \, .
\end{align}
For states of this form,  the quantum measurement that minimizes the error can be found with the method of Ref. \cite{pra2005}.  In the case at hand, the optimal measurement is covariant \cite{holevo} and is given by the operators
\begin{align}\label{Ameas}
M_h  =  |\eta_{h^{-1}}\>\<\eta_{h^-1}|  \qquad    |\eta_{h^{-1}}\>  :  =   \bigoplus_{k=0}^{2j}  (2k+1)     \,  |S_{k,h^{-1}}\> \, ,
\end{align}
with normalization   $\int \d h \,  M_h  =   I$. Here we have the inverse $h^{-1}$ (instead of just $h$, as in the usual definition of covariant measurement) because the scope of the measurement is  to rotate back the axes, resulting in the fact that the error $d(h,g)$ is minimum when $h=  g^{-1}$.   
  
Using the optimal measurement, the expected error is given by 
\begin{align}
\nonumber \left\<d^2\right\>   
   &  = \frac 13 \left(  4+2c_0^2 - 4 \sum_{k=1}^{2j}  c_k c_{k-1}  \right) \\
\nonumber  &  =  \frac 23 \,  [     c_{2j}  c_{2j+1}     -    c_0  c_1              +3c_0^2  \\
\label{Aerror}   &  \qquad   -    \sum_{k=1}^{2j}  c_k \left( c_{k+1}   +  c_{k-1}-  2 c_k\right)    ]  \, .
\end{align}

Now, let us assume also  that $c_k$ is an analytical function.
 Then, a third-order Taylor expansion   gives
\begin{align*}
c_{k+1}   +  c_{k-1}-  2 c_k     =   c^{(2)}  (k)   +    \frac 1 {4!}   \,    c^{(4)} \left (\xi_k^+\right)     +    \frac 1 {4!}       c^{(4)} \left (\xi_k^-\right)   \, ,
\end{align*}
where $c^{(l)}  (k)$ is the $l$-th derivative of $c_k$ and  $\xi_k^+$  ($\xi_k^- $)  is a point in    $[ k,k+1]$    ($[k-1, k]$).
Inserting this  expression in Eq.  (\ref{Aerror})  and using the Euler-MacLaurin formula,  we obtain
\begin{align}
\nonumber \left\<d^2\right\>  =&      \frac 23  \left\{     c_{2j}  c_{2j+1}     -    c_0  c_1            -  c_0^2  -     \int_{1}^{2j}   \d k ~ f(k)    \right.    \\
\nonumber   &  -  \frac 12 \left[   f(1)  +  f(2j)   \right]      \\
\nonumber  &  - \sum_{l=1}^p   \frac{B_{2l}}{(2l)!}    \left[  f^{(2l-1)}  (2j)    -     f^{(2l-1)}  (1)  \right]   +  R_p   \\
   & \left.  - \frac 1 {4!} \sum_{k=1}^{2j}   c_k  \left[  c^{(4)} \left (\xi_k^+\right) +  c^{(4)} \left (\xi_k^-\right)   \right]   \right\}\, ,
 \label{Aerror2}
\end{align}
where  the function $f$ is defined as $f(k)  :=  c_k   c^{(2)}(k)  $, $f^{(l)}$ is its $l$-th derivative, $B_l$ is the $l$-th Bernoulli number, and
\begin{align*}
|R_p|  \le \frac{2  \zeta (2p)}{(2\pi)^{2p}}  \int_{1}^{2j}    \d k    \left|f^{(2p)}(k)\right|  \, ,
\end{align*}
$\zeta$ being Riemann's  zeta function.

Going back to the specific case of two spin-$j$ singlets, in this case  $c_k  =  \sqrt{2k+1}/(2j+1)$.    We  use Eq. (\ref{Aerror2}) with $p=1$.        By direct inspection it is easy to see that the leading order  term is  $c_{2j} c_{2j+1}  = 1/j  +  O(1/j^2)$.
Indeed,  the integral in the fist line is of order $\ln j/j^2$, while all  the remaining terms are of order  $O(1/j^2)$ or higher.
For example, the sum in the last line of Eq. (\ref{Aerror2}) can be expressed as
\begin{align*}
\sum_{k=1}^{2j}c_k\left[  c^{(4)} \left (\xi_k^+\right) +  c^{(4)} \left (\xi_k^-\right)   \right]  =  \frac{-15}{(2j+1)^2}  \,      S_j  \, ,
\end{align*}
where
\begin{align*}
S_j  =  \sum_{k=1}^{2j}\sqrt{2k+1}\left[\left(2\xi^+_k+1
\right)^{-7/2}+\left(2\xi^-_k+1
\right)^{-7/2}\right] \, .
\end{align*}
is upper bounded by a constant.  Indeed, one has
\begin{align*}
S_j   \le   &2 \sum_{k=1}^{2j}\sqrt{\frac{2k+1}{2k-1}}\cdot(2k-1)^{-3}\\
  &  \le  2 \sqrt3\cdot
\sum_{k=0}^{\infty}(2k+1)^{-3}   \\
&  =     \frac   {  7 \sqrt 3 \,     \zeta  (3)   }{4} \, .
\end{align*}

In conclusion, we obtained the asymptotic expression $\left\<d^2\right\>  =  2/(3j)  +  O(\ln j/j^2)$.

\section*{Supplementary Note 2} 

Setting   $F_{\rm no}   =  \sqrt{  I  -  F_{\rm yes}^2}$, the state of the particles when the filter is not passed is of the form of Eq. (\ref{Astate}), with
\begin{align*}
c_k    =  \sqrt{  \frac{ (2k+1)/(2j+1)^2    -   p_{\rm yes}^{\rm opt}       \sin^2\left[\frac{\pi(k+1)}{2(j+1)}\right]/(j+1)    }{1  -  p_{\rm yes}^{\rm opt}}     }
\end{align*}
\begin{figure}
  \centering
  \includegraphics[width=0.9\linewidth]{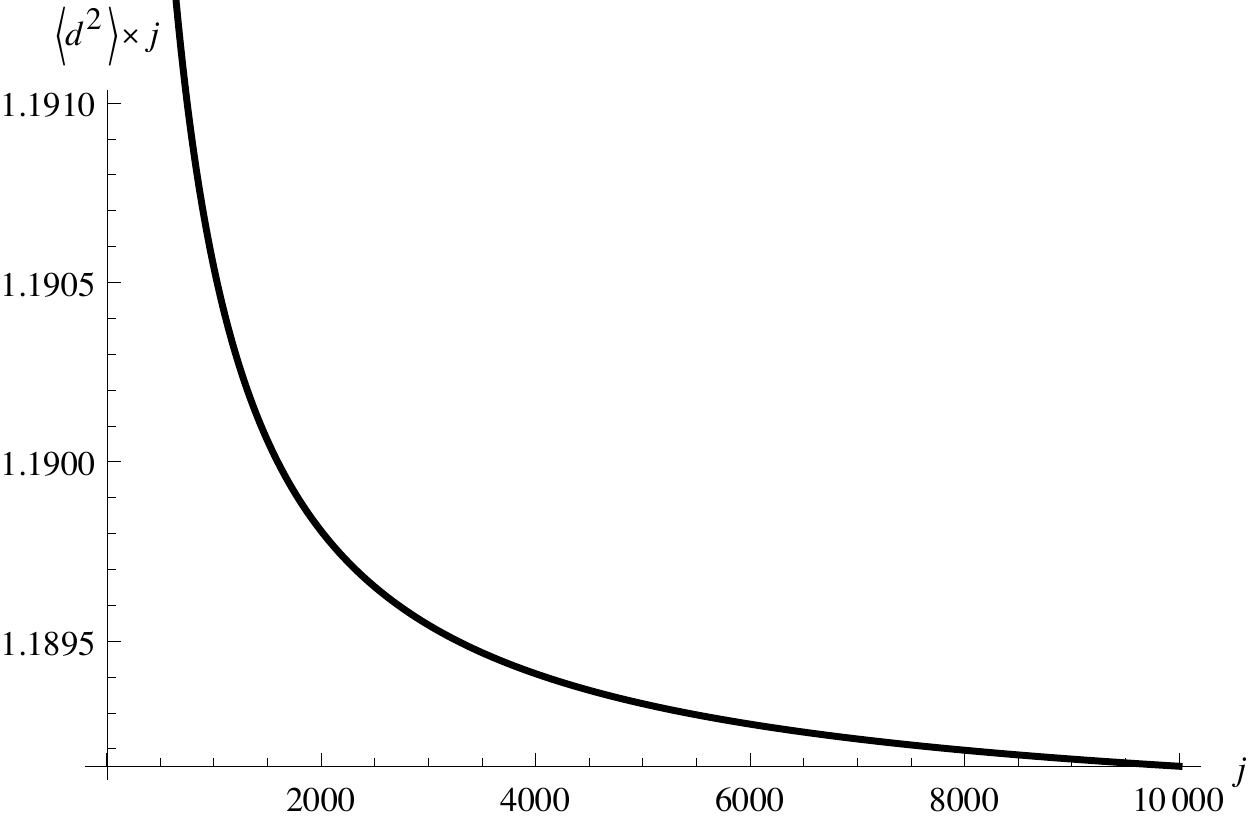}\\
  \caption{{\bf  SQL scaling for two spin-$j$ singlets  in the unfavourable case.}   The plot shows the values of the product  $\left\<d^2\right\> \times  j$ for two spin-$j$ singlets, conditioned on the  unfavourable outcome of the filter, with $j$ going from 10 to 10000 in unit steps.    Note the SQL scaling  $\left\<d^2\right\> \approx  1.189/j$ in the large $j$ limit. }\label{nogo}
   \end{figure}
Inserting this expression in Eq. (\ref{Aerror}) and evaluating the error we obtain the SQL scaling $\left\<d^2 \right\> \approx  1.189/j$  illustrated  in Figure \ref{nogo}.

\section*{Supplementary Note 3} 

\begin{figure}
  \centering
  \includegraphics[width=0.9\linewidth]{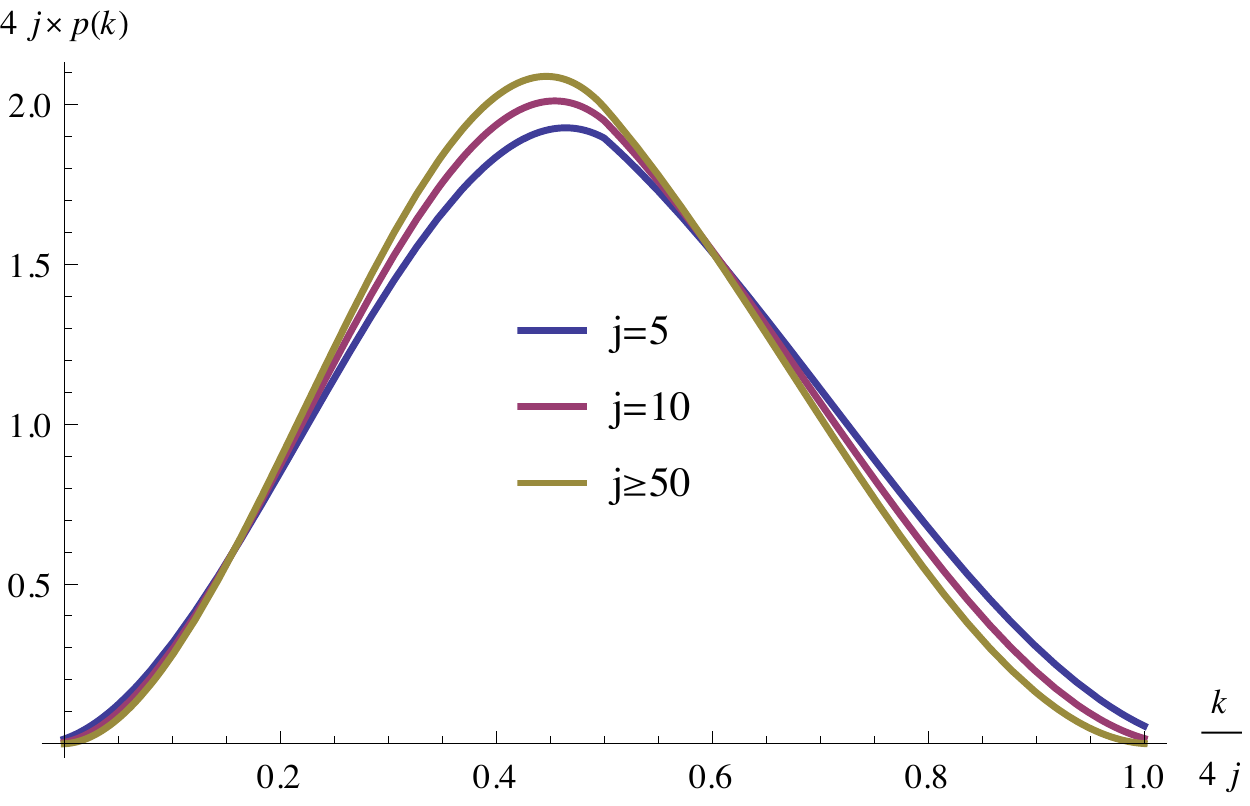}\\
 \caption{{\bf Probability distribution of the total angular momentum for  four spin-$j$ singlets.}  The plot is in suitably scaled  units, which guarantee the normalization of the total probability.  Practically, for $j\ge 50$ the distinction between the probability distribution and its limit for $j\to\infty$ is so small that it cannot be detected from the plot.}\label{fig:4copystate}
\end{figure}

The state  $|S_j\>^{\otimes4}$ is of the form of Eq. (\ref{Astate}) with
\begin{align*}     c_k  =     \frac{  \sqrt{(2k+1)  m_k  }  }  {(2j+1)^2}
\end{align*}
and
\begin{align*}
m_k  =
\left\{\begin{array}{ll}
-3k^2 /2  + 4 k j + k/2 + 2 j + 1   \qquad   & k  \le 2j   \\
 8 j^2 +k^2/2 -4kj+6j-3k/2 + 1 &  k>2j \, .
 \end{array}
 \right.
 \end{align*}
The probability distribution $p(k)  = c_k^2$  is shown in Figure  \ref{fig:4copystate} for different values of $j$. Note that, in suitable scaled units, the probability converges quickly to its limit value for $j\to \infty$.   

 Again, the error can be evaluated with Eq. (\ref{Aerror}). In order to obtain the asymptotic expression, we break the sum into two parts, one from $1$ to $2j$ and the other from $2j+1$ to $4j$, using Eq. (\ref{Aerror2}) to evaluate  the  first part and a similar expression for the second. A laborious  but straightforward calculation along the lines of  Supplementary Note 1  then shows that the leading contribution  to the error comes from the integral
 \begin{align*}
 I_4  &  = -\frac23\left( \int_{1}^{2j}     \d k \, c_k c^{(2)}(k)+\int_{2j+1}^{4j-1}     \d k \, c_k c^{(2)}(k) \right) \\
  &   =    11\ln(2)/ (18j^2)  + O(1/j^3) \, ,
  \end{align*}
  while the remaining terms are of order $O(1/j^3)$ or higher.   In conclusion, four spin-$j$ singlets allow one to achieve the HL scaling  $\left\<d^2\right\>   =      11\ln(2)/ (18j^2)  + O(1/j^3)$ with unit probability.

 \section*{Supplementary Note 4} 

The input state  $|S_j\>^{\otimes3}$ is of the form  of Eq. (\ref{Astate}),  with
\begin{align*}
c_k  =  \sqrt {  \frac{  (2k+1)  m_k    }  {(2j+1)^3}   }
\end{align*}
\begin{align*}
m_k  =
\left\{\begin{array}{ll}
  2k+1   \qquad   & k  \le j   \\
   3j+1-k &  k>j \, .
 \end{array}
 \right.
 \end{align*}

\begin{figure}
  \centering
  \includegraphics[width=0.9\linewidth]{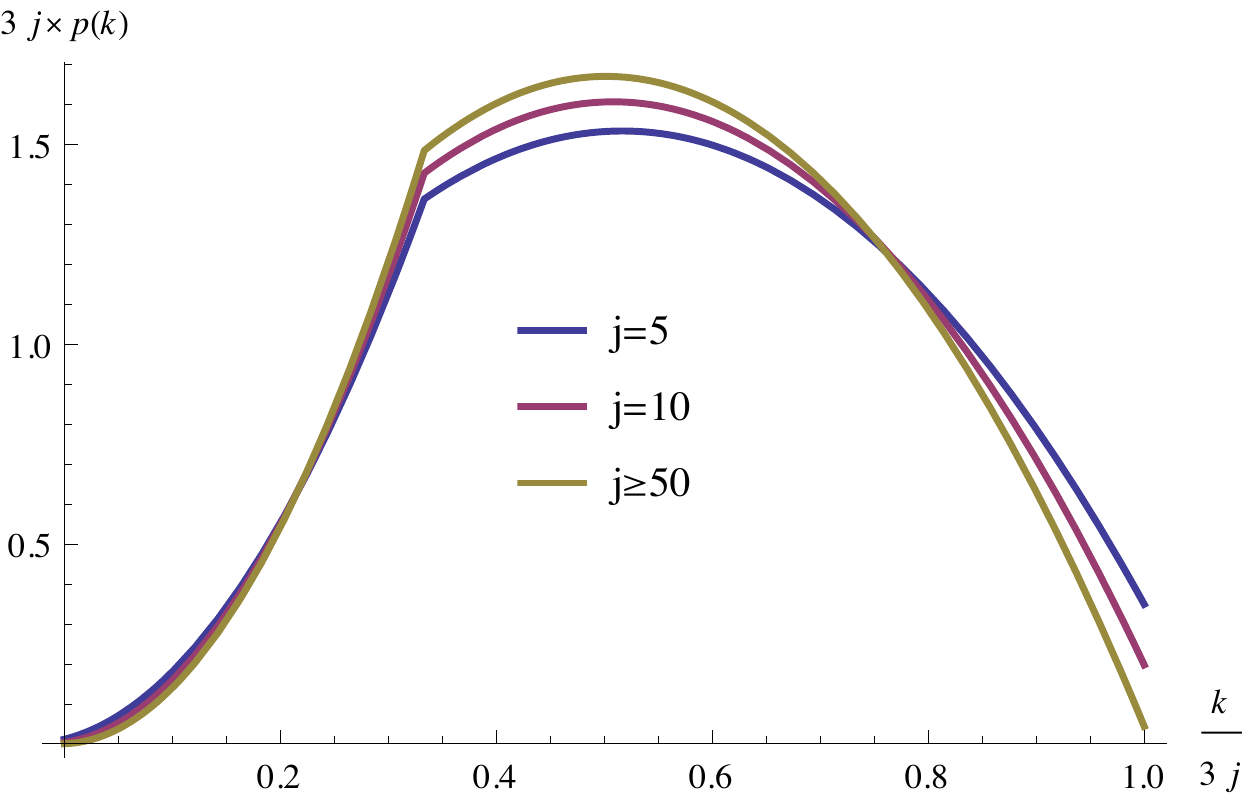}\\
  \caption{{\bf Probability distribution of the total angular momentum   for  three spin-$j$ singlets.}  The plot is in suitably scaled  units, which guarantee the normalization of the total probability.  Practically, for $j\ge 50$ the distinction between the probability distribution and its limit for $j\to\infty$ is so small that it cannot be detected from the plot.}  \label{fig:3copystate}
\end{figure}

The probability distribution $p(k)  = c_k^2$  is shown in Figure  \ref{fig:3copystate} for different values of $j$. Note that, again, the probability distribution in rescaled units converges quickly to its limit value for $j\to \infty$.

Now,  the error can be written as
\begin{align*}
 \left\<d^2\right\>
   &  = \frac 13   \, (  4+2c_0^2-4c_{3j}c_{3j-1}    \\
     & \qquad - 4 \sum_{k=1}^{j}  c_k c_{k-1}
    - 4 \sum_{k=j+1}^{3j-1}  c_k c_{k-1} \, )
\end{align*}

Following the same steps in Supplementary Note 1,     one can easily find that the leading contribution of the error comes from the integral \begin{align*}
I_3  &=-\frac23\int_{j+1}^{3j-1}\d k\,c_kc^{(2)}(k)  \\
&  =\ln j/(8j^2)+O(1/j^2) \, ,
\end{align*}
whereas all the remaining terms are of order $O(1/j^2)$ or higher.  In conclusion, three spin-$j$ singlets allow one to achieve the quasi-Heisenberg scaling $\left\<d^2\right\>  =  \ln j/(8j^2)+O(1/j^2)$.

 \section*{Supplementary Note 5} 

Let us we parametrize  the rotations as $U^{(j)}_g:=\exp\left[i\mathbf{J}^{(j)}\cdot\bs{\theta}\right],\ \mathbf{J}^{(j)}=(J^{(j)}_x,J^{(j)}_y,J^{(j)}_z)$ are the  angular momentum operators and $\bs  {  \theta}  =  (\theta_x,\theta_y,\theta_z)$ are real parameters.     Following Ref. \cite{QFI}, we find that the quantum Fisher information (QFI) matrix for the spin-$j$ singlet is given by
\begin{align*}
(F_Q)_{ik}&=4\left[\frac{1}{2}\<S_j| \left(J^{(j)}_iJ^{(j)}_k+J^{(j)}_kJ^{(j)}_i\right)\otimes I|S_j\>\right.\\
&~\left.  \phantom{\frac 12}-\<S_j|J^{(j)}_i\otimes I|S_j\>\<S_j|J^{(j)}_k\otimes I|S_j\>\right]\\
&=4j(j+1)\delta_{ik}/3,\qquad i,k=x,y,z \, .
\end{align*}
The quantum  CRB then becomes
\begin{align*}
 V_{\bs \theta}    \ge  F_Q^{-1}    \equiv    \frac{3}{4j(j+1)} \,  I   \, ,
\end{align*}
where $  V_{\bs \theta}$ is the covariance matrix of  $\bs \theta$ and $I$ is the $3\times 3$ identity matrix.

 \section*{Supplementary Note 6} 
We now fix $j$ and  analyze the asymptotic scaling of the error  for a large number  $n $ of  identical copies of a rotated  spin-$j$ singlet.   

In order to evaluate the error,  we express the state $  |S_{j,g}\>^{\otimes n}$ as
$|S_{j,g}\>^{\otimes n}  =    \bigoplus_{k  =  k_{\min}}^{nj}  \,  \sqrt{p^{(n)}_k}|S_{g,k}\>  $
where $k_{\min}   = 1/2$ if  $n$ is odd and $j$ is semi-integer and zero otherwise,  while   $p_n (k) $ is given by
\begin{align*}
p_{n,k}  =   (2k+1)   \,  \int \d g \,   \Tr\left[  U_g^{(k)} \right]  \,   \<  S_{j}  |   S_{j,g}  \>^n
\end{align*}

 Parametrizing the rotations in terms of the rotation angle (denoted by $\omega$) and of the the polar coordinates of the rotation axis (denoted by $\phi$ and $\theta$),
we obtain
\begin{align*}
p_{n,k}   &  =\frac{2k+1}{\pi}\int_{-\pi}^{\pi}\d\omega~\sin [(k+1/2)\omega  ]  \sin( \omega/2)   \\
   &  \qquad   \qquad  \qquad  \times   \,  \exp\left\{n\ln
\left[\frac{\sin(j+1/2)\omega}{(2j+1)\sin\frac{\omega}{2}}\right]\right\}  \\
&=\frac{2k+1}{\pi}\int_{-\pi}^{\pi}\d\omega~\sin [(k+1/2)\omega  ]  \sin( \omega/2)   \\
  & \qquad  \qquad  \times   \exp\left\{-\frac{nj(j+1)\omega^2}{6}-
  O(  n j^4 \omega ^4)   \right\} \\
& =  \frac{3\sqrt{3}(2k+1)^2}{2\sqrt{2\pi n^3j^3(j+1)^3}}\exp\left[-\frac{3k^2}{2nj(j+1)}\right]  \\
  &  \qquad  \qquad \qquad \qquad \qquad  \times \left[1-O\left(\frac{1}{n}\right)\right]  \, .
\end{align*}


Now,  the optimal quantum measurement \cite{pra2005} is given by the operators
\begin{align}\label{Ameas}
M_{g}  =  |\eta_{g^{-1}}\>\<\eta_{g^{-1}}|  \qquad    |\eta_{g^{-1}}\>  :  =   \bigoplus_{k=k_{\min}}^{nj}  (2k+1)     \,  |S_{k,g^{-1}}\> \, . 
\end{align}
Following the same arguments as in Supplementary Note 1 the  corresponding error can be expressed as
\begin{align*}
\nonumber \left\<d^2\right\>  =&      \frac 23  \left\{     c_{nj}  c_{nj+1}     -    c_{k_{\min}}  c_{k_{\min}  + 1}            -  c_{k_{\min}}^2  -     \int_{k_{\min} + 1}^{nj}   \d k ~ f(k)    \right.    \\
\nonumber   &  -  \frac 12 \left[   f(k_{\min} +1)  +  f(nj)   \right]      \\
\nonumber  &  -  \frac{B_{2}}{2}    \left[  f^{(1)}  (nj)    -     f^{(1)}  (k_{\min} + 1)  \right]   +  R   \\
   & \left.  - \frac 1 {4!} \sum_{k=k_{\min} +1}^{nj}   c_k  \left[  c^{(4)} \left (\xi_k^+\right) +  c^{(4)} \left (\xi_k^-\right)   \right]   \right\}\, ,
 \end{align*}
with  $c_k=\sqrt{p_{n,k}}$, $f(k)  =   c_k  c_k^{(2)}$ and $ \xi_k^+   \,  (\xi_k^-)$ is a point in $[ k, k+1]  \,  (  [  k-1, k])$.   It is now easy to check that the leading terms is given by  the integral
\begin{align*}
I_3  &=-\frac23\int_{k_{\min} + 1}^{nj}\d k\,  f(k)  \\
&  = \frac{3}  {  2nj(j+1)  } +O  \left( \max \left\{n^{-3/2}  j^{-3}  ,    n^{-2}  j^{-2}   \right   \}  \right)
\end{align*}
whereas all the remaining terms are of order $O\left (   n^{-3/2}   j^{-3}   \right)$   or higher.   In conclusion, we obtained the asymptotic scaling
\begin{align}\label{asymptotic}
\left\<d^2\right\>  =    \frac{3}   {2nj(j+1)}     +   O \left( \max \left \{    n^{-3/2}  j^{-3}, n^{-2}  j^{-2}  \right \}     \right ) \, .
\end{align}

\medskip

 \section*{Supplementary Note 7} 
We now show that the optimal covariant measurement of  Eq. (\ref{Ameas}) attains the quantum CRB for large $n$.   Note that this is not a priori clear, since
 the measurement (\ref{Ameas}) is optimal for the minimization of the error---but may not be optimal in the CRB sense.

To show optimality, we first note that the CRB implies directly a  lower bound on the error, which can be easily evaluated in the parametrization $g =  g(\bs \theta)$. Indeed, using a       Taylor expansion to the second order we get    \begin{align*}
d^2(g,e)  
  =   \frac{2}{3}    ( \theta_x^2  +   \theta_y^2  +  \theta_z^2)  +O(\|\bs{\theta}\|^3),
\end{align*}
which, averaged  over $\bs \theta$, becomes
\begin{align}\label{Aerrorfinal}
\left\<  d^2\right\>      =  \frac  23   \Tr[   V_{\bs \theta}]  +    O( n^{-3/2}  j^{-3}  )  \, .
\end{align}
In bounding the error term,  we  exploited   the fact that, due to Eq. (\ref{asymptotic}) and Chebyshev's inequality,   the probability distribution of the outcomes of the optimal measurement is concentrated in a neighborhood of size $O\left(   n^{-1/2}  j^{-1}\right)$ centred around the identity.

Now, the QFI for   $n$ copies is given by  $F_Q^{(n)}   =  n  F_Q  =   4 n  j(j+1)  \, I/3   $.     Inserting the quantum CRB in the r.h.s. of  Eq. (\ref{Aerrorfinal}),   we  obtain the bound
\begin{align*} \left\<d^2\right\>\ge    \frac 3 {2n j(j+1)}   +       O\left( n^{-3/2}  j^{-3} \right )   \,.
\end{align*}
By comparison with Eq. (\ref{asymptotic}) we conclude that the optimal measurement for $n$ copies satisfies $ \Tr\left  [  V^{\rm opt}_{{\bs \theta},n}\right]   =    4n  j(j+1)  +   O \left( \max \left \{    n^{-3/2}  j^{-3}, n^{-2}  j^{-2}  \right \}     \right ) $.   Since the optimal measurement (\ref{Ameas}) satisfies also   \begin{align*}
\<  \theta_i \theta_j\>   =  \delta_{ij}   \frac  { \Tr \left [  V^{\rm opt}_{{\bs \theta},n}\right]} 3 \, ,
\end{align*}
we conclude that for this measurement one has 
\begin{align*}  V^{\rm opt}_{{\bs \theta},n}  =          \frac{3}{4nj(j+1)} \,  I      +   O \left( \max \left \{    n^{-3/2}  j^{-3}, n^{-2}  j^{-2}  \right \}     \right )    \, .
\end{align*}   In other words, the  optimal covariant measurement achieves the quantum CRB for large $n$.

\section*{Supplementary Note 8} 

Let us consider an arbitrary separable measurement, of the form $M_{h}   =  \sum_i  A_{i,h}  \otimes B_{i,h}$ for positive operators $A_{i,h}$ and $B_{i,h}$.    Using a standard averaging argument \cite{holevo}, one can show that the optimal measurement can be chosen to be covariant,~i.~e.~of the form $ M_h  = ( \map  U_{h^{-1}}  \otimes  \map I_B) ( M_e )$. 

Now, the normalization of the measurement gives
\begin{align*}
I_A\otimes I_{B}  &  = \int  \d h\,  M_h \\
  &  =  \sum_i    \Tr  [  A_i]  \,     \frac {  I_A}{2j+1}  \otimes   B_i  \, ,
\end{align*}
having used the irreducibility of the representation. Redefining $A_i'   : =   (2j+1)  A_i $ and $  B_i'  =  B_i/(2j+1)$ we then obtain
\begin{align*}
\sum_i   B_i'    =  I_B   \qquad {\rm and}  \qquad  \int \d h  \,  \map U_h  (  A_i')     =  I_A  \, .
\end{align*}
Hence, every separable measurement can be realized as a one-way LOCC measurement, where Bob performs the POVM  $\{ B_i'\}$ and communicates the outcome to Alice, who performs the POVM $\{A^{(i)}_h\}$ with operators $A^{(i)}_h  :  =  \map U_h  (A_i')$.    In such a protocol, Alice has to perform the optimal POVM for the conditional state induced by Bob's measurement.  The error that can be achieved in this way  is lower bounded by the error in the estimation of $h$ from the state $|\psi_h\>=  U_h  |\psi\>$, where $ |\psi\> $ is the best input state for a spin-$j$ particle.  To evaluate this lower bound, we can use  a covariant POVM $\{  M_h\}$.  In this case, covariance implies that $M_h$ is of the form $M_h  = (2j+1) \map U_h  (\rho)$ for some quantum state $\rho$ \cite{holevo}. By convexity of the figure of merit, the optimal POVM has $\rho  =  |\psi'\>\<\psi'|$ for some pure state.   Hence, the error is
\begin{align*}
\<d^2\>   &   =   (2j+1) \min_{\psi,\psi'}   \int \d h  \,    e( h, e)  \,    |\<  \psi  | U_h |\psi'\>|^2  \\
  &  =     2 \left[    1   -  \frac {2j+1} 9       \max_{\psi, \psi'}    \< \psi  |  \<  \tilde \psi|     \Pi_1  |\psi'\>  |\tilde \psi'\>   \right]  \\
    & \ge        2 \left[    1   -  \frac {2j+1} 9       \max_{\psi}    \< \psi  |  \<  \tilde \psi|     \Pi_1  |\psi\>  |\tilde \psi\>   \right] \\
  &   =  (2j+1) \min_{\psi}    \int \d h  \,    e( h, e)  \,  |\<  \psi  |  U_h  |\psi\>|^2     \, ,
\end{align*}
where  $\Pi_1$ is the projector on the eigenspace with total quantum number $j=1$ and $ |\tilde \psi\>  :  =    e^{  i\pi  J_y}    |\bar \psi\>$ with $|\bar \psi\>$ the complex conjugate of $|\psi\>$.       For $j= 1/2$, all pure states are equivalent under rotations and therefore there is no need  of further optimization.      Plugging $|\psi\>  =  |1/2,1/2\>$ in the equation we obtain the value  $\left\<d^2\right\>  =  16/9  $.

\end{document}